\documentclass[twocolumn,showpacs,prl,amsmath,amssymb]{revtex4}
\usepackage{graphicx}     % Include figure files
\begin{document}
 
\title{Theoretical Evidence for Equivalence between 
the Ground States of the Strong-Coupling BCS Hamiltonian and 
the Antiferromagnetic Heisenberg Model}
\author{K. Park}
\affiliation{Condensed Matter Theory Center, 
Department of Physics, University of Maryland,
College Park, MD 20742-4111}
\date{\today}

\begin{abstract}
By explicitly computing wavefunction overlap via exact diagonalization
in finite systems,
we provide evidence indicating that, 
in the limit of strong coupling, i.e., $\Delta/t \rightarrow \infty$,
the ground state of the Gutzwiller-projected BCS Hamiltonian
(accompanied by proper particle-number projection)  
is identical to the exact ground state of 
the 2D antiferromagnetic Heisenberg model on the square lattice. 
This identity is adiabatically connected to
a very high overlap between the ground states of 
the projected BCS Hamiltonian and the $t$-$J$ model 
at moderate doping. 
\end{abstract}

\maketitle
%%%%%%%%%%%%%%%%%%%%%%%%%%%%%%%%%%%%%%%%%%%%%%%%%%%%%%%%%%%%%%%%%%%%%%%%%%

One of the key questions regarding high $T_C$ superconductivity
is whether it is possible for a repulsive electron-electron interaction
alone to give rise to superconductivity. 
%let alone the high $T_C$.
Considering that  
(i) the Hubbard model can be transformed into 
the $t$-$J$ model in the limit of large on-site repulsion,
and (ii), at half filling, the $t$-$J$ model reduces to 
the two-dimensional (2D) $S=1/2$ antiferromagnetic Heisenberg model
with nearest-neighbor exchange coupling on the square lattice, 
the answer to the above question seems to lie in
the relationship between
2D quantum antiferromagnetism 
and superconductivity.
Although there is a consensus that the ground state of
the 2D antiferromagnetic Heisenberg model
has N\'{e}el order,
i.e., staggered spin order \cite{DR},
the precise nature of the ground state wavefunction itself, 
especially upon doping,
is controversial.

The resolution of controversy requires 
both an {\it unbiased} and {\it unambiguous} study. 
%It is difficult to satisfy both criteria of 
%unambiguity and unbiasedness.
Many analytic approaches are based on uncontrolled approximations
such as the large-$N$ expansion \cite{Fradkin}.
Even numerical approaches such as 
variational Monte Carlo simulation \cite{Gros,PRT}
are biased by 
the choice of trial state  
which is assumed to be the ground state.
%Without the independent verification of trial state,
%it is not clear how relevant the trial state is 
%to a specific microscopic model.
Exact diagonalization, 
on the other hand,
is a completely unbiased approach. 
%in attacking strongly correlated systems. 
However, its application is limited to finite systems
with relatively small spatial size. 
%of 4-6 lateral lattice spacings.
This limitation generates a problem since 
the evidence of an ordered state is usually given by 
correlation functions.
In the case of superconductivity,
the relevant correlation function is 
the pairing correlation function:
$F_{\alpha\beta}({\bf r}-{\bf r}') = \langle 
c^{\dagger}_{\uparrow}({\bf r})
c^{\dagger}_{\downarrow}({\bf r}+{\bf \alpha})
c_{\downarrow}({\bf r}')
c_{\uparrow}({\bf r}'+{\bf \beta}) \rangle$
where ${\bf \alpha}, {\bf \beta} = \hat{x}, \hat{y}$.
Off-diagonal long-range order can be claimed only when
$F_{\alpha\beta}$ remains
non-zero in the limit of large $|{\bf r}-{\bf r}'|$.
Unfortunately,
small spatial sizes of finite systems accessible via
exact diagonalization make 
the distinction between
true long-range order and 
short-range order (present even in normal states) ambiguous.

Despite the small spatial size, however,
the size of Hilbert space is quite large;
depending on the number of holes,
there are $10^3$-$10^5$ basis states 
in the $4\times4$ lattice system
(even after translational symmetries are implemented
as reported in this paper). 
Since the ground state is constructed out of
a very large number of possible linear combinations of basis states,
it is a highly non-trivial task to infer a good ansatz
wavefunction for the ground state itself.
%even if the system size is relatively small in real space.
Consequently, if there is a method to show directly that
some form of the BCS wavefunction is identical to 
the exact ground state of 
the 2D antiferromagnetic Heisenberg model,
it will 
provide convincing evidence 
for superconductivity upon doping.
In this paper, we put forward such a method: 
we compute the wavefunction overlap between
the ground states of the $t$-$J$ model and
the Gutzwiller-projected BCS Hamiltonian
via exact diagonalization.
%we compute the wavefunction overlap between 
%the exact ground state (obtained from exact diagonalization)
%and the Gutzwiller-projected BCS wavefunction 
%(with proper particle-number projection). 
It should be emphasized that, in this paper,
the Gutzwiller projection is applied directly
in the BCS Hamiltonian instead of being applied
onto the BCS ground state wavefunction.
Note that the Gutzwiller-projected BCS wavefunction
was originally proposed by Anderson \cite{Anderson}
as a candidate for the ground state 
either on the square lattice 
with sufficiently strong next-nearest-neighbor exchange coupling, 
or on the triangular lattice.
The Gutzwiller-projected BCS wavefunction
was then conjectured to be a good ground state for
the $t$-$J$ model on the square lattice at moderate, non-zero doping. 
The difference between the ground state of 
the Gutzwiller-projected BCS Hamiltonian
and the Gutzwiller-projected BCS ground state
is most crucial at half-filling, which
will be discussed in detail later in this paper.

The inspiration for using wavefunction overlap
comes from the fractional quantum Hall effect (FQHE).
It is well accepted by now that all essential aspects
of FQHE are explained by 
the composite fermion (CF) theory \cite{Jain}.
%which has received strong support 
%from both experiment and theory.          
%However, 
One of the main reasons why we can put unequivocal trust 
in the CF theory may be the amazing agreement between
the exact ground state and the CF wavefunction: 
the overlap is practically unity
for various short-range interactions 
including the Coulomb interaction \cite{Perspectives}.
In this paper, we would like to achieve the same methodological clarity
for the $t$-$J$ model
crucial in establishing the CF theory.

Before we discuss computational details,
it is illuminating to note that
the analogy with FQHE goes much deeper than just methodology.
%In fact, the way in which the FQHE problem was solved
%can serve as an enlightening physical guide in attacking
%general strongly correlated problems.
To gain a physical insight into
why the CF wavefunction is so accurate,
let us consider the Laughlin wavefunction \cite{Laughlin}
which is a subset of CF wavefunctions 
at special lowest-Landau-level filling factor $\nu= 1/(2p+1)$ 
with $p$ an integer; 
especially at $\nu=1/3$, it is given by
$\Psi_{1/3}=\prod_{i<j} (z_i -z_j)^3$ where $z=x+iy$.
Trugman and Kivelson \cite{TK}, 
and Haldane \cite{HaldanePseudo} showed
that $\Psi_{1/3}$ is actually
the exact ground state of a short-range interaction
given by $\nabla^2 \delta({\bf r})$.
%The proof is basically as follows.
%Any 2D orbital state of two electrons interacting via
%isotropic potential can be decomposed into 
%each relative angular momentum ($L_z$) channel.
%The potential strength in each $L_z$ channel,
%called the Haldane pseudopotential in the FQHE,
%determines the ground state wavefunction which
%optimizes amplitudes in each $L_z$ channel
%to minimize the total energy cost.
%It can be shown \cite{HaldanePseudo} 
%that $V_{\textrm{TK}}({\bf r})$ 
%is repulsive only in the $L_z =1$ channel,
%and vanishes in all other channels of odd $L_z$
%(note that only odd-$L_z$ channels matter for fully spin polarized electrons).
%The functional form of $\Psi_{1/3}$ shows that
%every electron pair in $\Psi_{1/3}$
%consists exclusively in the $L_z=3$ channel so that
%it completely avoids any interaction energy cost.
%While it is encouraging that $\Psi_{1/3}$ 
%is the exact ground state of $V_{\textrm{TK}}({\bf r})$,
%it is still not guaranteed that the Laughlin state
%is also a good representation 
%of the ground state of Coulomb interaction
%which is more relevant for experiments.
%So, it is crucial to verify how close 
%the Laughlin state is to the exact ground state.
%As mentioned previously, this is accomplished
%by computing the wavefunction overlap 
%which is basically unity.
For the Coulomb interaction (relevant for experiments),
the Laughlin state remains extremely close to
the exact ground state, 
as shown by practically unity overlap.
%Physically, the important lesson is 
%that the Jastrow-type correlation in $\Psi_{1/3}$
%is very effective in minimizing interaction energies.
The breakthrough for the CF theory
was achieved when Jain realized that 
the Laughlin state is actually composed of two parts:
the Jastrow factor, $\prod_{i<j} (z_i -z_j)^2$
(dubbed flux quanta attachment 
because of concomitant phase winding),
and a non-interacting fermionic wavefunction 
of new quasiparicles, i.e., composite fermions.
%Jain then proposed the CF ansatz wavefunction 
%for general filling factors. 
%which is the product of Jastrow factor and non-interacting CF wavefunction 
%at appropriate effective filling factors.
%Unlike the Laughlin state, the general CF state
%may not be generated from
%a simple interaction in a closed form.
%However, its overlap with exact states is so close to unity
%in all known cases that there is little doubt 
%in the validity of CF wavefunction.
The key physical point is that,
once the short-range correlation is captured
by the Jastrow factor, 
residual correlations
can be treated as relatively weak long-range correlations
which are much easier to handle.

%Actually, there is another well-known strongly correlated system, 
%for which the most accurate microscopic theory is formulated 
%in terms of the variational ansatz wavefunction. 
%That is the normal liquid $^3$He \cite{Mahan}.
%The ground state of normal liquid $^3$He 
%is given by an algebraic product of the Bijl-Dingle-Jastrow factor
%and a Slater determinant of plane-wave states (Fermi sea state):
%$\Psi_{^3\textrm{He}} = 
%\exp{[-\sum_{i<j} u({\bf r}_i-{\bf r}_j)]}
%\textrm{Det}|e^{i{\bf k}\cdot{\bf r}}|$
%where $u({\bf r})$ is determined by 
%the interaction between He atoms.
%While the Slater determinant provides the necessary antisymmetrization
%and the low-energy, long-range physics, 
%the Jastrow factor provides
%the necessary short-range correlation.

%We have shown so far that separating out the short-range correlation
%in terms of the Jastrow factor played an important role
%in studies of various strongly correlated systems. 
%We believe, in fact, that this approach can be taken as a general method
%in solving strongly correlated systems.
In fact, 
the separation between 
short-range and long-range correlations
can serve as a general method in attacking
strongly-correlated problems.
The main question is
what type of short-range correlation exists 
in the specific problem at hand, 
and more importantly
what functional form of Jastrow factor describes 
this short-range correlation.
Since we are interested in the quantum antiferromagnetism, 
%and its connection to high $T_c$ superconductivity,
it is natural to ask what has been known conclusively 
in the context of short-range correlation 
in quantum antiferromagnetic models.
%and its implementation in an ansatz wavefunction approach.

To this end, let us consider the 1D $S=1/2$ 
quantum Heisenberg model. 
This model is important because
its solution is exactly known for two important cases of
nearest-neighbor and $1/r^2$ exchange coupling.
For the nearest-neighbor exchange coupling,
the Bethe ansatz solution \cite{Bethe} gives the exact ground state
which, for a given spin configuration,
has an amplitude equal to the product of plane-wave states
of spin flip excitation with appropriate phase shift.
%determined by two-magnon scattering process.
%Physically speaking, the Bethe ansatz solution is possible
%because, in one dimension,
%two-magnon scattering process conserves individual momenta, 
%regardless of forward or backward scattering
%which differ only by a phse shift.
Since the Bethe ansatz solution is basically
a product of plane-wave states,
it is encouraging to guess that some form of Fermi sea state 
might be closely related to the exact Bethe ansatz solution,
which turns out to be precisely the case.
Various numerical works \cite{KHF,GJR}
as well as exact analytic studies \cite{GV}
showed that, in addition to closeness in energy, 
the spin-spin correlation function of
the Gutzwiller-projected Fermi sea state 
has a power-law behavior very similar to the exact result. 
Note that the Gutzwiller projection simply imposes 
the no-double-occupancy constraint.
Furthermore, Haldane \cite{Haldane1D} and Shastry \cite{Shastry}
proved that
the Gutzwiller-projected Fermi sea state is the exact ground state
of the 1D $S=1/2$ Heisenberg model with $1/r^2$ exchange coupling.

Now, combined with the fact that 
the Gutzwiller projection is basically an implementation of 
strong on-site repulsion,
the above-mentioned similarity leads to a conjecture that
the Gutzwiller projection plays the role of
a Jastrow factor providing 
the short-range correlation 
embedded in quantum antiferromagnetism. 
To support this,
we compute the overlap 
between the ground states of 
the Gutzwiller-projected BCS Hamiltonian 
and the antiferromagnetic Heisenberg model
%we compute the overlap 
%between the Gutzwiller-projected BCS state and 
%the exact ground state of the 2D Heisenberg model
(in general, the $t$-$J$ model at non-zero doping).
In fact, we will show that the ground state of
the projected BCS Hamiltonian
is identical to the exact ground state 
of the Heisenberg model
in the limit of strong coupling.
%In fact, we will show that the projected BCS state
%is identical to the exact state 
%in the limit of strong pairing.
Also, the overlap is very high ($\sim 90 \%$)
in a realistic parameter range relevant to cuprates,
which is adiabatically connected to 
the unity overlap in the aforementioned limit.

We begin our quantitative analysis by writing the Hamiltonian
of the $t$-$J$ model:
\begin{eqnarray}
H_{t\textrm{-}J} &=& \hat{{\cal P}}_G ( H_t + H_J ) \hat{{\cal P}}_G ,
\nonumber \\
H_t &=& -t \sum_{\langle i,j \rangle} 
(c^{\dagger}_{i\sigma} c_{j\sigma} + \textrm{h.c.}),
\nonumber \\
H_J &=& J \sum_{\langle i,j \rangle} 
({\bf S}_i \cdot {\bf S}_j -n_i n_j /4),
\label{H_tJ}
\end{eqnarray}
where $\hat{{\cal P}}_G$ is the Gutzwiller projection operator
imposing the no-double-occupancy constraint.
We obtain the exact ground state of $H_{t\textrm{-}J}$
using a modified Lanczos method.
We have checked that our results for the $t$-$J$ model 
completely agree with previous numerical studies \cite{Dagotto}
for all available cases.

Now, let us turn our attention to 
the Gutzwiller-projected BCS Hamiltonian.
As mentioned in the beginning, 
our approach is rather different from previous approaches
%Our method of constructing the Gutzwiller-projected BCS state
%is rather different from previous approaches 
\cite{Gros,HP,PRT,Sorella}
which applied the number projection as well as
the Gutzwiller projection onto 
the explicit BCS wavefunction.
We do not take these previous approaches for two reasons.
First, there is a singularity due to 
the spin-density-wave instability
inherent in the BCS Hamiltonian with on-site repulsion,
which generates N\'{e}el order
at half filling. So, it is important to study
directly the Gutzwiller-projected BCS Hamiltonian
instead of applying the Gutzwiller projection
onto the BCS ground state.
The Gutzwiller-projected BCS Hamiltonian is given as follows:
%First, we are interested in computing the overlap with
%the exact state obtained from numerical diagonalization.
%So, it is far more convenient to obtain the Gutzwiller-projected BCS state
%directly in terms of the same basis states 
%used in numerical diagonalization.
%So, we consider the following Hamiltonian 
%(represented in real space coordinates), 
%for which the BCS state is the exact ground state:
\begin{eqnarray}
H_{\textrm{BCS}} &=& H_t + H_{\Delta} , 
\nonumber \\
H_{\Delta} &=& \Delta \sum_i 
(c^{\dagger}_{i\uparrow} c^{\dagger}_{i+\hat{x},\downarrow}
-c^{\dagger}_{i\downarrow} c^{\dagger}_{i+\hat{x},\uparrow}
+\textrm{h.c.}) 
\nonumber \\
&-&\Delta \sum_i 
(c^{\dagger}_{i\uparrow} c^{\dagger}_{i+\hat{y},\downarrow}
-c^{\dagger}_{i\downarrow} c^{\dagger}_{i+\hat{y},\uparrow}
+\textrm{h.c.}) ,
\nonumber \\
H^\textrm{G}_{\textrm{BCS}} &=& 
\hat{{\cal P}}_G H_{\textrm{BCS}} \hat{{\cal P}}_G ,
\label{H_BCS}
\end{eqnarray}
where $H_t$ is given in Eq.(\ref{H_tJ}).
Note the sign change in front of $\Delta$  
in the $y$ direction compared to that of $x$ direction:
%With help of the Bogoliubov transformation,
%it is straightforward to solve $H_{\textrm{BCS}}$
%to obtain the BCS state with $d$-wave paring symmetry: 
the gap function is given in momentum space
by $\tilde{\Delta}({\bf k})=2\Delta(\cos{k_x}-\cos{k_y})$.
%while the kinetic energy is given by 
%$\epsilon({\bf k})=-2t(\cos{k_x}+\cos{k_y})$.
%We obtain the Gutzwiller-projected BCS state 
%by diagonalizing $H^\textrm{G}_{\textrm{BCS}}$.
%In other words, 
The Gutzwiller projection is built in
from the onset by working solely in the Hilbert space
with the no-double-occupancy constraint.
%which is the exactly the same Hilbert space where
%the $t$-$J$ model is diagonalized. 
%Then, we perform the number projection which, 
%in turn,  leads to
%the second reason for our approach.

The second reason is rather subtle, but physically very important.
Coherent number fluctuations 
(as opposed to incoherent fluctuations in the normal state)
are ultimately responsible for 
the intrinsic properties of superconductivity. 
%such as
%the Josephson effect as well as zero resistance itself.
So, coherent number fluctuations should
be incorporated 
even into finite-system studies
in a fundamental manner.
In essence, we diagonalize $H^\textrm{G}_{\textrm{BCS}}$
in the combined Hilbert space of $N_e$ and $N_e+2$ particles.
%because coherent number fluctuations
%and associated pairing correlations are fully captured
%within this combined Hilbert space.
There is, however, a spurious finite-size effect 
which prevents pairing 
if one naively  
diagonalizes $H^\textrm{G}_{\textrm{BCS}}$
in the combined Hilbert space.
In finite systems, the energy cost of adding (removing) few particles 
is not negligible compared to the total energy.
So, the mixing between states 
with even a few-particle difference  
is energetically prohibited.
We fix this problem by adjusting the chemical potential
so that the kinetic energy plus the chemical potential energy
of the $N_e$ particle ground state is the same as that of the 
$N_e+2$ particle ground state, which eliminates a spurious energy 
penalty for pairing.  
Once the chemical potential is set, it can be shown that 
the mixing with other particle-number sectors such as the 
$N_e+4$ sector is negligibly small, even if it is allowed.
%Note that the issue of chemical potential 
%in the context of superconductivity in finite system
%is not properly addressed in 
%previous studies \cite{Gros,HP,PRT,Sorella}.

%%%%%%%%%%%%%%%%%%%%%%%%%%%%%%%%%%%%%%%%%%%%%%%%%%%%%%%%%%%%%%%%%%%%%%%%%
\begin{figure}
\includegraphics[width=3.5in]{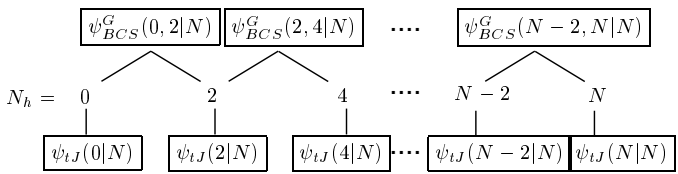}
\caption{Schematic diagram for constructing
the ground state of the Gutzwiller-projected BCS Hamiltonian, 
$\psi^{G}_{BCS}(N_h,N_h + 2|N)$,
and the exact ground state of the $t$-$J$ model, 
$\psi_{t\textrm{-}J}(N_h|N)$.
$N_h$ ($N$) is the number of holes (sites).
\label{fig1}}
\end{figure}
%%%%%%%%%%%%%%%%%%%%%%%%%%%%%%%%%%%%%%%%%%%%%%%%%%%%%%%%%%%%%%%%%%%%%%%%%%
Let us define the following notations:
$\psi^{G}_{BCS}(N_h,N_h + 2|N)$ denotes 
the ground state of the Gutzwiller-projected BCS Hamiltonian
obtained from
the combined Hilbert space of $N_h$ and $N_h + 2$ holes
in the system of $N$ sites.
$\hat{{\cal P}}_{N_h=N_0}$ denotes 
the number projection operator which projects
states onto the Hilbert space of $N_0$ holes
and re-normalizes the projected states.
$\psi_{t\textrm{-}J}(N_h|N)$ is the exact ground state of 
the $t$-$J$ model
in the Hilbert space of $N_h$ holes in $N$ sites.
A schematic diagram is shown in Fig.\ref{fig1}.

We now present our exact-diagonalization results of overlap
for various numbers of holes 
in the $4\times4$ square lattice system
with periodic boundary conditions. 
%\cite{comment_2by2}.
Note that the $4\times4$ system is one of the most studied
systems in numerics \cite{DR}
because it is accessible
via exact diagonalization, 
yet large enough to contain 
essential many-body correlations.
While it is possible to study all possible dopings 
in the $4\times4$ system,
we concentrate on the three most important cases:
0 hole (undoped regime),
2 holes (optimally doped regime), and
4 holes (overdoped regime).
Also, we study only an even number of holes since,
in finite systems,
an odd number of holes will artificially frustrate pairing order.

%%%%%%%%%%%%%%%%%%%%%%%%%%%%%%%%%%%%%%%%%%%%%%%%%%%%%%%%%%%%%%%%%%%%%%%%%
\begin{figure}
\includegraphics[width=1.8in,angle=-90]{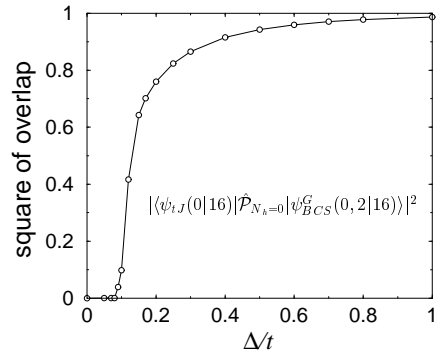}
\caption{Square of the overlap between the 
ground state of the projected BCS Hamiltonian,
$\hat{{\cal P}}_{N_h=0}|\psi^G_{BCS}(0,2|16)\rangle$,
and the exact ground state of the $t$-$J$ model,
$|\psi_{t\textrm{-}J}(0|16)\rangle$,
for the 0-hole case in the $4\times4$ system (undoped regime).
\label{fig2}}
\end{figure}
%%%%%%%%%%%%%%%%%%%%%%%%%%%%%%%%%%%%%%%%%%%%%%%%%%%%%%%%%%%%%%%%%%%%%%%%%%
Let us begin with the 0-hole case, i.e., 
the 2D antiferromagnetic Heisenberg model.
Though the numerical evidence for N\'{e}el order 
is quite convincing \cite{DR},
our knowledge of the ground state itself
is very limited,
considering that 
(i) the semiclassical configuration of staggered spins 
is not the exact ground state,
and (ii) there are still rather strong quantum fluctuations.
So, it will be satisfactory if one can show that the ground state 
of the Gutzwiller-projected BCS Hamiltonian is a good representation 
of the exact ground state of the 2D antiferromagnetic Heisenberg 
model. The criterion for the effectiveness of the ground state of the 
projected BCS Hamiltonian is quantified via its overlap with the 
exact ground state of the Heisenberg model. High overlap will 
provide evidence for the existence of pairing 
which, in turn, generates superconductivity upon doping.

When Anderson \cite{Anderson} proposed the RVB state 
(which is a synonym for the projected BCS state),
his insight was that electrons are already paired at zero doping,
but the ground state cannot superconduct 
(for that matter, conduct) 
because there are no mobile charge carriers. 
It seems reasonable, then, that the ground state 
becomes superconducting
as soon as holes are added.
However, the idea of the RVB state 
as the ground state of the Heisenberg model
was rejected because
the RVB state does not have any long-range magnetic order, 
while the exact ground state has N\'{e}el order \cite{CHN,RY}.
The situation is quite different 
for the ground state of the Gutzwiller-projected 
BCS Hamiltonian.
%This notion is incorrect because 
%magnetic properties of projected BCS state 
%depend on $\Delta/t$.
Fig.\ref{fig2} shows that, at zero doping,
the ground state of the projected BCS Hamiltonian,
$\hat{{\cal P}}_{N_h=0}|\psi^G_{BCS}(0,2|16)\rangle$,
is actually identical to 
the exact ground state of the $t$-$J$ model,
$|\psi_{t\textrm{-}J}(0|16)\rangle$,
in the limit of $\Delta/t \rightarrow \infty$.
Since the two states are identical,
it is clear that the ground state of the projected BCS Hamiltonian
has N\'{e}el order.
This identity at infinite $\Delta/t$
is consistent with previous Monte Carlo simulations \cite{PRT},
in which their variational gap parameter 
becomes very large at small doping.
%In hindsight,
%the reason why previous studies \cite{Gros} incorrectly
%concluded that the RVB state had no long-range magnetic order
%was that their parameter range was wrong,
%i.e., $\Delta/t$ was set finite.
It is important to know that
the largeness of $\Delta/t$ does not necessarily mean
strong superconductivity since, 
despite strong pairing,
there is very little charge fluctuation 
at small doping \cite{comment_F}.
%As mentioned in the beginning,
%the true measure of ODLRO is
%$F_{\alpha\beta}({\bf r}-{\bf r}')$ in the limit of 
%large $|{\bf r}-{\bf r}'|$,
%which can be quite generally shown 
%to vanish as the square of hole concentration 
%at small dopings 
%for the projected BCS state \cite{PRT}.

But, physically, why is the overlap unity
at infinite $\Delta/t$, or equivalently $t=0$?
%When $t=0$, electrons cannot move, 
%and therefore there is no charge fluctuation.
%Charge fluctuations are naturally suppressed in 
%the $t$-$J$ model because, at zero doping,
%$H_t$ has vanishing matrix elements regardless of $t$. 
%In $H_{BCS}$, however, 
%there are intrinsic charge fluctuations due to pairing.
%Lowering $t$ in $H_{BCS}$ should suppress charge fluctuations
%yielding the same physics as the $t$-$J$ model.
To answer this,
consider the $t$-$J$ Hamiltonian at zero doping, $H_J$,
and the BCS Hamiltonian with $t=0$, $H_\Delta$.
$H_J$ contains ${\bf S}_i\cdot{\bf S}_j$
which prefers the formation of spin singlet pairs.
Similarly, $H_\Delta$ contains
$c^{\dagger}_{i\uparrow}c^{\dagger}_{j\downarrow}
-c^{\dagger}_{i\downarrow}c^{\dagger}_{j\uparrow}$
which also prefers to create spin singlet pairs.
So, provided that 
the no-double-occupancy constraint is imposed 
via Gutzwiller projection,
$H_J$ and $H_\Delta$ should have the same physical effect.

%%%%%%%%%%%%%%%%%%%%%%%%%%%%%%%%%%%%%%%%%%%%%%%%%%%%%%%%%%%%%%%%%%%%%%%%%
\begin{figure}
\includegraphics[width=2.8in]{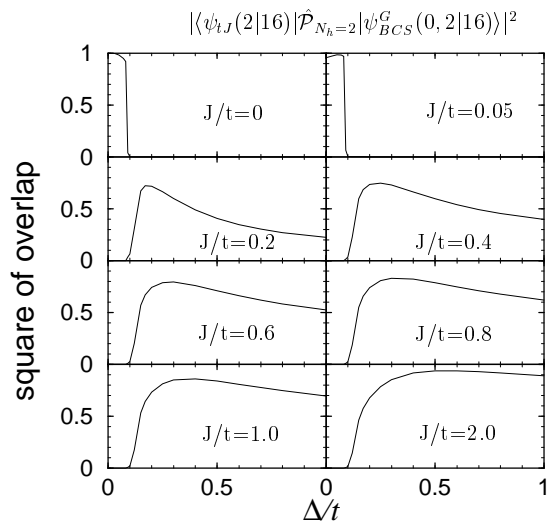}
\caption{Square of the overlap between the ground state of 
the projected BCS Hamiltonian,
$\hat{{\cal P}}_{N_h=2}|\psi^G_{BCS}(0,2|16)\rangle$,
and the exact ground state of the $t$-$J$ model,
$|\psi_{t\textrm{-}J}(2|16)\rangle$,
for the 2-hole case in the $4\times4$ system 
(optimally doped regime).
\label{fig3}}
\end{figure}
%%%%%%%%%%%%%%%%%%%%%%%%%%%%%%%%%%%%%%%%%%%%%%%%%%%%%%%%%%%%%%%%%%%%%%%%%%
We now move onto the cases with finite doping.
Since pairing already exists at zero doping,
it is expected that
the ground state becomes superconducting upon doping. 
We support this by showing that,
in the Hilbert space of two holes,
the ground state of the projected BCS Hamiltonian,
$\hat{{\cal P}}_{N_h=2}|\psi^{G}_{BCS}(0,2|16)\rangle$,
has a very high overlap ($\sim 90\%$) 
with $|\psi_{t\textrm{-}J}(2|16)\rangle$
at optimal $\Delta/t$ 
for a realistic range of $J/t$: 
$0.4 \lesssim J/t \lesssim 0.8$ 
(middle panels in Fig.\ref{fig3} showing the square of the overlap).
Note that $\Delta/t$ can be taken as a variational paramter.
In fact, the optimal overlap approaches unity
when $J/t$ bocomes sufficiently large 
(bottom, right in Fig.\ref{fig3}).
While the large $J/t$ regime itself is not very realistic,
the high overlap in the realistic regime 
is adiabatically connected to 
the unity overlap in the large $J/t$ limit.
Incidentally,
the identity between $\psi_{t\textrm{-}J}$ and $\psi^{G}_{BCS}$ 
at $J/t=0$ and $\Delta/t=0$ 
(top, left in Fig.\ref{fig3}) 
is rather trivial because $H_{t\textrm{-}J}$ and $H^{G}_{BCS}$
become identical in this case.
It is important to note that the nature of the identity at large $J/t$
is completely different from that of zero $J/t$, as manifested 
by symmetry changes of the ground state.
The $t$-$J$ model ground state  
changes its rotational symmetry from $s$-wave to $d$-wave
at $J/t \simeq 0.08$ 
while the ground state of the projected BCS Hamiltonian
does so at $\Delta/t \simeq 0.1$.
Therefore, the regime with large $J/t$ and $\Delta/t$ is
completely disconnected from the regime with small $J/t$ and $\Delta/t$.
It is important to distinguish the rotational symmetry 
of the ground state from the pairing symmetry. The latter 
is always $d$-wave while the former changes as a function 
of $\Delta/t$.

Finally, we have checked that, in the overdoped regime,
the ground state of the projected BCS Hamiltonian
is no longer a good representation of 
the ground state of the $t$-$J$ model,
which is supported by the negligible overlap between
$\hat{{\cal P}}_{N_h=4}|\psi^G_{BCS}(2,4|16)\rangle$
and $|\psi_{t\textrm{-}J}(4|16)\rangle$
for general parameter range.

In conclusion, we have provided evidence that, in the limit of 
strong coupling, the ground state of the Gutzwiller-projected BCS 
Hamiltonian is equivalent to that of the 2D antiferromagnetic 
Heisenberg model. Combined with high overlaps at moderate doping,
this equivalence supports the existence of superconductivity 
in the $t$-$J$ model. For future work, it will be interesting to 
investigate an analytic approach in proving the equivalence.

The author is very grateful to S. Das Sarma for his support 
throughout this work, and to J. K. Jain for his insightful comment 
on the difference between the projected Hamiltonian and the 
projected ground state. The author is also indebted to V. Galitski,
D. J. Priour, Jr. and V. W. Scarola for valuable discussions. 
This work was supported by ARDA and NSF.

%%%%%%%%%%%%%%%%%%%%%%%%%%%%%%%%%%%%%%%%%%%%%%%%%%%%%%%%%%%%%%%%%%%%%%%

%%%%%%%%%%%%%%%%%%%%%%%%%%%%%%%%%%%%%%%%%%%%%%%%%%%%%%%%%%%%%%%%%%%%%%%
\end{document}